\documentclass[aps,pre,preprint,groupedaddress,showpacs]{revtex4}

\usepackage{graphicx}
\usepackage{color}
\usepackage{amsmath}
\usepackage{amssymb}

\def\be{\begin{equation}}
\def\en{\end{equation}}
\def\bea{\begin{eqnarray}}
\def\ena{\end{eqnarray}}

\def\n{\nabla}

\newcommand{\av}[1]{\langle{#1}\rangle}

\newcommand{\bi}[1]{\mbox{\boldmath$#1$}}
\newcommand{\pp}[2]{\frac{\partial {#1}}{\partial {#2}}}

\begin{document}

\title{Polydomain growth at isotropic-nematic 
transitions in liquid crystalline polymers}

\author{Shunsuke Yabunaka and Takeaki Araki}
\affiliation{Department of Physics, Kyoto University, 
Sakyo-ku, Kyoto 606-8502, Japan}

\date{\today}

\begin{abstract}
We studied the dynamics of isotropic-nematic 
transitions in liquid crystalline polymers 
by integrating time-dependent Ginzburg-Landau 
equations. 
In a concentrated solution of rodlike polymers, 
the rotational diffusion constant $D_{\rm r}$ of the polymer 
is severely suppressed 
by the geometrical constraints of the surrounding polymers, 
so that the 
rodlike molecules
diffuse only along their rod directions. 
In the early stage of phase transition, the 
rodlike polymers with nearly parallel orientations
assemble to form a nematic polydomain. 
This polydomain pattern with 
characteristic 
length $\ell$, grows with self-similarity in three dimensions (3D)
over time with a $\ell\sim t^{1/4}$ scaling law.
In the late stage, the rotational diffusion 
becomes significant,
leading a crossover of the growth exponent from 
$1/4$ to $1/2$. 
This crossover time is estimated to be of 
the order $t\sim D_{\rm r}^{-1}$. 
We also examined time evolution 
of a pair of disclinations 
placed in a confined system, 
by solving the same time-dependent Ginzburg-Landau equations in two dimensions (2D).
If the initial distance between the disclinations is 
shorter than some critical length, 
they approach and annihilate each other; 
however, at larger initial separations they are stabilized.

\end{abstract}

\maketitle

\section{Introduction}

Liquid crystalline polymers (LCPs) 
are widely used in technology as high-performance fibers 
since they have high strength and are light in weight. 
Their mechanical strength increases 
when the high molecular weight polymers are orientated in the same direction 
\cite{Doi-Edwards,Marrucci1989,See1990,Marrucci1993,
Larson1991,Shimada1988b,Shimada1988c}. 
Compared to low molecular weight liquid crystals (LMWLCs), 
LCP molecules do not align spontaneously.
In this paper, we study the polydomain 
formation after quenching an LCP system 
from the isotropic to the nematic state. 
The phase transition dynamics in LCPs have been 
mostly studied on the basis of the conventional 
nematohydrodynamic equations \cite{Feng2001,Klein2007,Fu2008}, 
which were developed 
to describe the dynamics of LMWLCs \cite{deGennes}. 
We are thus interested in the possible differences 
in the pattern evolution in LCPs and LMWLCs. 

In a concentrated solution, 
rodlike polymers entangle with one another, and hence, the 
surrounding polymers strongly suppress 
the rotational and perpendicular diffusions \cite{Doi-Edwards}. 
In the high concentration limit, 
each rodlike polymer moves only along its molecular axis and 
this parallel diffusion dominates the dynamics 
of the isotropic-nematic transition.

If the rotational motion of the director is absent, 
the orientational order parameter behaves as a conserved variable \cite{Shimada1988c}. 
It is well known that the mechanism of domain growth and the
resultant growth exponent depend on whether 
its order parameter is preserved. 
For a system described by a single non-conserved order 
parameters such as magnetization, 
the domain pattern with a characteristic length $\ell$ 
grows in time as $\ell\propto t^{1/2}$ \cite{Onukibook}. 
On the contrary, when the order parameter is preserved, 
the domain growth obeys $\ell\propto t^{1/3}$ as observed 
in the phase separation of 
binary mixtures \cite{Onukibook}. 
The former is termed as ``model A" and the latter 
as ``model B" \cite{Hohenberg1977}.

In a typical LMWLC, each molecule can freely rotate to 
align parallel to the surrounding molecules. 
The phase transition dynamics are well described by the time-dependent 
Ginzburg-Landau equation with a non-conserved tensorial order 
parameter \cite{deGennes}. 
In the late stage, 
the characteristic length of the polydomain pattern $\ell$ 
grows in time as $\ell\propto t^{1/2}$ 
\cite{Toyoki1994, Fukuda1998, Denniston2001}. 
For LCPs, 
Shimada {\it et al.} studied the early stage of the 
phase transition and predicted a spinodal decomposition of 
the nematic order parameter \cite{Shimada1988c}. 
However, since their kinetic equations are 
linearized, the domain growth in the late stage 
could not be treated. 
In this study, 
we reformulate 
the free-energy functional and kinetic equations for rodlike polymers 
to include non-linear terms. 
This enables us to analyze the late stage behaviors 
of the phase ordering, 
such as domain growth and defect motions. 
Similar kinetic models for mixtures of isotropic liquids 
and semi-flexible polymers 
have been proposed by several authors \cite{Liu1996,Fukuda1999}. 
However, these studies have focused on the phase separation in the mixtures and 
isotropic-nematic transitions have not yet been studied. 
The main aim of this paper is to elucidate 
the isotropic-nematic 
transition dynamics in concentrated solutions of LCPs 
and to determine the effects of small but finite polymer rotational diffusions on the system.

This article is organized as follows: 
In section 2 we formulate 
free-energy functional and kinetic equations for the rodlike polymers 
in accordance with the method of Shimada {\it et al.} 
In section 3, we show the results of the numerical 
simulations on the isotropic-nematic transitions and discuss them. 
In section 4, we summarize our work.

\section{Free-energy functional and kinetic equations}

Similar to Shimada {\it et al.} \cite{Shimada1988b,Shimada1988c}, 
we derive the free-energy functional and kinetic equations for 
a solution of LCPs. 
We reformulate their linearized equations to time-dependent 
Ginzburg-Landau equations \cite{Hohenberg1977} in order to study 
the late stage of the phase transition. 

Rodlike polymers of length $d$ and width $w$ ($w\ll d$) are considered. 
For a solution of the rodlike polymers, 
we introduce the free-energy functional 
$\mathcal{F}$ 
for 
$f({\bi r},{\bi u})$, 
which represents 
the probability 
distribution of rods at position $\bi{r}$ where 
$\bi{u}$ is 
the unit vector along the rod direction. 
It consists of two parts as shown in Eq. (\ref{eq:free_ene}) 
\bea
&&\mathcal{F}=\mathcal{F}_0+\mathcal{F}_{\rm int}.
\label{eq:free_ene}
\ena
$\mathcal{F}_0$ is the free-energy functional for ideal 
non-interacting polymers and is expressed as 
\bea
\mathcal{F}_0
=k_{\rm B}T\int d\bi{r} d\bi{u}f({\bi r},{\bi u}) 
\{ \ln (v_0f({\bi r},{\bi u}))-1\}, 
\label{eq:F0}
\ena
where $T$ is the temperature, $k_{\rm B}$ is the Boltzmann constant, and
$v_{0}(\cong w^2d)$ is the volume of the rodlike polymer. 
The second term in Eq. (\ref{eq:free_ene}) 
is the interaction part of 
the free-energy functional and is expressed as 
\bea
&&\mathcal{F}_{\rm int}=\frac{1}{2}k_{\rm B}T 
\int d\bi{r}d\bi{u}\int d\bi{r}'\bi{u}'
W(\bi{r},\bi{u},\bi{r}',\bi{u}')\nonumber\\
&& \hspace{1cm} \times f(\bi{r},\bi{u})
f(\bi{r}',\bi{u}').
\label{eq:interaction} 
\ena
$W(\bi{r},\bi{u},\bi{r}',\bi{u}')$ represents the excluded volume interaction 
between two rodlike polymers $(\bi{r},\bi{u})$ and $(\bi{r}',\bi{u}')$, and is defined by 
\begin{eqnarray}
W(\bi{r},\bi{u},\bi{r}',\bi{u}')=\left\{ \begin{array}{ll}
\alpha & (\mbox{when two polymers intersect}) \\
0 & (\mbox{otherwise}), \\
\end{array} \right. 
\end{eqnarray}
where $\alpha$ is the interaction parameter that has the dimension of volume and is estimated to be $wd^2$ 
\cite{Straley1973,Shimada1988b}.
We neglect the interaction between the polymers and solvent; therefore, the isotropic-nematic transition originates purely 
from the configurational 
entropy of the rods \cite{Onsager,deGennes}. 
In other words, 
the solution is an athermal system, in which temperature 
changes play no role in the phase behaviors. 
Above a critical concentration, the solution exhibits 
a liquid crystalline phase. 
This free-energy functional is applicable not only to LCP solutions 
but also to suspensions of rigid rods, 
such as the tobacco mosaic virus (TMV) 
\cite{Zasadzinski1986,Graf1999,Urakami1999} 
and carbon nanotubes \cite{Zhang2006}.

We define two order parameters as 
\bea
\phi(\bi{r})&=&v_0\int d\bi{u} f(\bi{r},\bi{u}),\\
Q_{ij}(\bi{r})&=&v_0\int d\bi{u} f(\bi{r},\bi{u})
\left(u_iu_j-\frac{1}{3}\delta_{ij}\right), 
\ena
where $\int d\bi{u}$ represents a solid angle integration. 
$\phi$ is the concentration of the polymers, 
and $Q_{ij}$ is the orientational order per volume. 
It is noted that $Q_{ij}$ vanishes with vanishing $\phi$, 
even when the polymers are orientationally ordered. 
Although it is more natural to use $Q_{ij}/\phi$ as 
an order parameter, 
we evaluate the free-energy functional with $\phi$ and $Q_{ij}$ 
for simplicity.

The distribution function can be expanded for $\phi$ 
and $Q_{ij}$ as shown in Eq. (\ref{eq:f})
\bea
f(\bi{r},\bi{u})=\frac{1}{4\pi v_0}\left\{\phi(\bi{r})
+\frac{15}{2}Q_{ij}(\bi{r})
\left(u_iu_j-\frac{1}{3}\delta_{ij}\right)\right\}. 
\label{eq:f}
\ena
Hereafter, the repeated suffixes $i$ 
and $j$ indicate summation over 
$i,j=x,y,z$. 
We substitute Eq. (\ref{eq:f}) into Eq. (\ref{eq:F0}), 
and integrate it only over $\bi{u}$ using the 
isotropic approximation (see Appendix A). 
After some calculations, 
the free energy for ideal polymers 
$\mathcal{F}_0$ 
is expanded 
to include the fourth order of $Q_{ij}$ as shown in Eq. (\ref{eq:F0_2}) 
\bea
&&\mathcal{F}_0
=\frac{k_{\rm B}T}{v_0}\int d\bi{r}\left\{\phi\ln\frac{\phi}{4\pi e}
+\frac{a_0}{2\phi} Q_{ij}Q_{ji}
-\frac{b}{3\phi^2}Q_{ij}Q_{jk}Q_{ki}
\nonumber\right.\\
&&\hspace{1.5cm}\left.+\frac{c}{4\phi^3}(Q_{ij}Q_{ji})^2
+\frac{c'}{4\phi^3} Q_{ij}Q_{jk}Q_{kl}Q_{li}\right\},
\label{eq:F0_2}
\ena
where the numerical constants are $a_0=15/2$, $b=225/14$, 
$c=375/28$, and $c'=375/7$.

Using the isotropic approximation, 
we obtain the interaction part of the free energy 
[Eq. (\ref{eq:interaction})] 
in reciprocal $\bi{q}$-space as follows: 
\bea
&&\mathcal{F}_{\rm int}=\frac{\pi k_{\rm B}T\alpha}{{v_{0}}} 
\int \frac{d\bi{q}}{(2\pi)^3}\Biggl[
\left(\frac{1}{4}-\frac{\bi{q}^2d^2}{144}
+\frac{11\bi{q}^4d^4}{115200}\right)|\tilde{\phi}(\bi{q})|^2\nonumber\\
&&+\frac{7}{768}d^2q_iq_j \tilde{\phi}(\bi{q}) \tilde{Q}_{ij}(-\bi{q})
-\frac{15}{64}\tilde{Q}_{ij}(\bi{q})\tilde{Q}_{ji}(-\bi{q})\nonumber\\
&&+\frac{15}{5376}d^2 \{q_kq_k\tilde{Q}_{ij}(\bi{q})\tilde{Q}_{ij}(-\bi{q})
+4|q_j\tilde{Q}_{ij}(\bi{q})|^2\}\Biggr], 
\label{eq:free_q}
\ena 
where $\tilde{X}({\bi{q}})$ is the Fourier $\bi{q}$-component of 
the variable $X(\bi{r})$ in reciprocal space. 
The expression of $W$ in reciprocal space is denoted in Appendix B. 
With reverse Fourier transformation, we finally obtain 
$\mathcal{F}_{\rm int}$ in real space as 
\bea
\mathcal{F}_{\rm int}&=&\frac{k_{\rm B}T}{v_0}\int d\bi{r}\Biggl\{
\frac{\chi}{2} \phi^2-\frac{a_1}{2}Q_{ij}Q_{ji}\nonumber\\
&&+\frac{C_0}{2}|\nabla\phi|^2+\frac{C_1}{2}|\n^2\phi|^2
+K_0\nabla_i\phi \n_j Q_{ij}\nonumber\\
&&+\frac{K_1}{2}\left|\n_iQ_{jk}\right|^2+\frac{K_2}{2}
\left|\n_iQ_{ij}\right|^2\Biggr\},
\label{eq:free_ene2}
\ena
where 
$\chi=\pi\alpha/(2v_0)$, $a_1=15\pi\alpha/(32v_0)$, 
$C_0=-\pi \alpha d^2/(72v_0)$, $C_1=11\pi\alpha d^4/(57600v_0)$,
$K_0=7\pi \alpha d^2/(768v_0)$, 
$K_1=15\pi \alpha d^2/(2688v_0)$, $K_2=15\pi \alpha d^2/(672v_0)$, and 
$\n_i$ represents $\partial/\partial r_i$ $(i=x,y,z)$. 
In our model, 
the system is in the 
isotropic state for $\phi<\phi_1(=976v_0/63\pi\alpha)$ and 
in the nematic state for $\phi>\phi_2(=16v_0/\pi\alpha)$. 
When $\phi_1<\phi<\phi_2$, both the phases coexist. 

It is noted that the gradient terms of $\phi$ are expanded 
up to the fourth order of $\bi{q}$ because the coefficient 
of $\bi{q}^2|\tilde{\phi}|^2$ is negative in Eq. (\ref{eq:free_q}). 
This negative coefficient implies 
that the density modulations have a periodicity of $2\pi(-C_1/C_0)^{1/2}$, 
and its contribution on the phase ordering is small 
as discussed in Appendix C. 
However, actually, our numerical simulations do not show such a modulated pattern 
in our concentration range.

Next, we introduce the auxiliary fields $\mu$ and $H_{ij}$ as 
\bea
\frac{\delta\mathcal{F}}{\delta f}=v_{0}
\left\{\mu+H_{ij}\left(u_iu_j-\frac{1}{3}\delta_{ij}\right)\right\}.
\label{eq:mu_H}
\ena
With isotropic approximation, the variation of the free-energy 
functional is given by 
\bea
\delta \mathcal{F}=\int d\bi{r}(\delta \phi \mu +\delta Q_{ij}H_{ij}).
\ena
From this equation, the following expressions are derived, 
\bea
\mu=\frac{\delta \mathcal{F}}{\delta \phi},\quad H_{ij}
=\frac{\delta\mathcal{F}}{\delta Q_{ij}}, 
\ena
in which $\mu$ and $H_{ij}$ can be interpreted as 
the chemical potential of $\phi$ and the molecular 
force field of $Q_{ij}$, respectively. 
We should note that 
our definition of $H_{ij}$ is different from the 
conventional definition in terms of its sign.

Using the free-energy functional $\mathcal{F}$, 
the Fokker-Planck equation (see Eq. (2.1) of Ref. \cite{Shimada1988c}) 
is rewritten as 
\bea
\pp{}{t}f=\frac{\bar{\phi}}{k_{\rm B}T}
\left[\n\cdot\left\{D_\parallel \bi{u}\bi{u}
+D_\perp({\rm \bf I}-\bi{u}\bi{u})\right\}
\n\frac{\delta \mathcal{F}}{\delta f}
+D_{\rm r}\mathcal{R}^2\frac{\delta \mathcal{F}}{\delta f}
\right]. 
\label{eq:Fokker-Planck2}
\ena
Here, 
$\bar{\phi}$ is the average concentration of the rodlike molecules and 
$\mathcal{R}=\bi{u}\times (\partial/\partial \bi{u})$ is 
the rotation operator \cite{Doi-Edwards}. 
$D_{\parallel}$ and $D_{\perp}$ 
are diffusion constants for 
parallel and perpendicular motions to the 
rod direction $\bi{u}$, respectively, and 
$D_{\rm r}$ is the coefficient for the 
rotational motion. 
For a dilute solution, the Kirkwood theory estimates 
the diffusion constants as 
$D_{\parallel}^0=k_{\rm B}T\ln\left(d/w\right)/(2\pi\eta_{\rm s}d)$, 
$D_{\perp}^0=k_{\rm B}T\ln(d/w)/(4\pi\eta_{\rm s}d)$, and 
$D_{\rm r}^0=3k_{\rm B}T\{ \ln\left(d/w\right)-\gamma\}
/(\pi\eta_{\rm s}d^{3})$. 
Here, $\eta_{\rm s}$ is the solvent viscosity and 
$\gamma$ is Euler's constant \cite{Doi-Edwards}. 
As already noted, in the high concentration limit, 
each rodlike polymer moves only along its molecular axis, 
namely, $D_{\parallel}\gg D_{\rm r}d^{2}\,,D_{\perp}$. 
Hereafter, we use the kinetic coefficients 
$L_X=D_X\bar{\phi}v_0/k_{\rm B}T$, 
where $X$ represents $\parallel\,,\perp$, and ${\rm r}$.

From Eq. (\ref{eq:Fokker-Planck2}), we express 
the kinetic equations for $\phi$ and $Q_{ij}$ with isotropic approximations as follows
\bea
&&\pp{}{t}\phi=\left(\frac{1}{3}L_\parallel+\frac{2}{3}L_\perp\right)\n^2\mu
+\frac{2}{15}
(L_\parallel-L_\perp)\n_i\n_jH_{ij},
\label{eq:develop_phi2}\\
&&\pp{}{t}Q_{ij}=\frac{2}{15}(L_\parallel-L_\perp)
\left(\n_i\n_j-\frac{1}{3}\n^2\delta_{ij}\right)\mu
+\frac{2(L_\parallel-L_\perp)}{105\nonumber}\\
&&\times \left\{\n^2H_{ij}+2(\n_i\n_kH_{kj}+\n_j\n_kH_{ki})
-\frac{4}{3}\delta_{ij}\n_k\n_lH_{kl}\right\}\nonumber\\
&&+\frac{2}{15}L_\perp \n^2H_{ij}-\frac{4}{5}L_{\rm r}H_{ij}.
\label{eq:develop_Q2}
\ena
The off-diagonal coefficients in Eqs.(\ref{eq:develop_phi2}) and 
(\ref{eq:develop_Q2}) 
satisfy the Onsager's reciprocal relationship, and in Eq. (\ref{eq:develop_phi2}), we omitted $\n^2H_{ij}\delta_{ij}/3$. 
The time derivative of the free-energy functional is given by 
\bea
&&\frac{d}{dt}\mathcal{F}=\int d\bi{r}\left(\mu \pp{\phi}{t}+H_{ij}\pp{Q_{ij}}{t}\right)
\nonumber\\
&&=-\int d\bi{r}
\left[\left(\frac{1}{3}L_\parallel+\frac{2}{3}L_\perp\right)(\n_i\mu)^2
\right.\nonumber\\
&&+\frac{4}{15}(L_\parallel-L_\perp)(\n_i\mu)(\n_jH_{ji})
+\frac{8}{105}(L_\parallel-L_\perp)(\n_jH_{ji})^2\nonumber\\
&&\left.+\left(\frac{2}{105}L_\parallel+\frac{4}{21}L_\perp\right)(\n_kH_{ij})^2+\frac{4}{5}L_{\rm r}(H_{ij})^2\right].
\label{eq:F_t}
\ena
In the second line of Eq. (\ref{eq:F_t}), 
we ignore the influence of $\phi$ and $Q_{ij}$ from outside the system. 
The conditions $L_\parallel\ge L_\perp \ge 0$ and $L_{\rm r}\ge 0$ 
guarantee that 
the integrand of Eq. (\ref{eq:F_t}) is positive, resulting 
in $d\mathcal{F}/dt\le 0$.

We normalize space and time by 
$d$ and $t_0=d^2/(D_\parallel\bar{\phi})$. 
In an aqueous suspension of TMV, we estimate that $d=320$ nm and $w=20$ nm. 
Assuming $T=300$ K and $\eta_{\rm s}=0.9\, {\rm mPa\cdot s}$, 
the diffusion constant is $D_\parallel\cong 6.3\, \mu{\rm m}^2/{\rm s}$, 
and we thus obtain $t_0\cong 102.4\bar{\phi}^{-1}$ ms. 
We integrate the coupled equations in the lattice space 
with the explicit Euler method. 
In order to save computational costs, 
the spatial $\Delta x$ and temporal $\Delta t$ increments are
varied according to the state point. 
In all the following simulations, we employ periodic boundary conditions.

\section{Results and discussion}

\subsection{Spinodal decomposition}
\label{sec:spinodal}

First, we study the isotropic-nematic transition in a 
concentrated solution of rodlike polymers. 
We perform 3D simulations and 
set $\alpha=16 v_0$, which corresponds to $d/w=16$. 
The spatial and temporal increments are 
$\Delta x=0.4d$ and $\Delta t=0.016 t_0$, respectively. 
As an initial condition, we set $\bar{\phi}=0.5$, 
which is larger than the critical concentration $\phi_2$. 
The spatial averages of all the components of $Q_{ij}$ are set 
to zero and we add random noise to them.
Here, we set $L_\perp=L_{\rm r}d^2=0$ in Eqs. (\ref{eq:develop_phi2}) 
and (\ref{eq:develop_Q2}), so that the phase transition proceeds in the rodlike polymers only via 
the diffusions 
along their axes. 

Shimada {\it et al.} reported spinodal 
decomposition-like growth of the nematic order parameter 
in the early stage of phase ordering \cite{Shimada1988c}. 
They also claimed that the domain growth can be 
separated 
into three modes, {\it i.e.,}, splay, twist, and bend. 
Accordingly, we decompose the structure factor 
of $Q_{ij}$ into 
\bea
S_{\rm spl}(\bi{q})&=& |a_i \tilde{Q}_{ij}a_j|^2, 
\label{eq:S_spl}\\
S_{\rm tws}(\bi{q})&=& 2|a_i \tilde{Q}_{ij}b_j|^2+2|a_i 
\tilde{Q}_{ij}c_j|^2, 
\label{eq:S_tws}\\
S_{\rm bnd}(\bi{q})&=& |b_i \tilde{Q}_{ij}b_j-c_i 
\tilde{Q}_{ij}c_j|^2+4|b_i \tilde{Q}_{ij}c_j|^2
\label{eq:S_bnd}. 
\ena
Here, $\bi{a}(=\bi{q}/|\bi{q}|)$ is the unit vector 
toward the wave vector $\bi{q}$, and $\bi{b}$ and $\bi{c}$ 
are also unit vectors, which are orthogonal to $\bi{a}$ and each other. 

\begin{figure}[htbp]
\begin{center}
\includegraphics[angle=0,width=0.5\textwidth, bb=0 0 553 394]{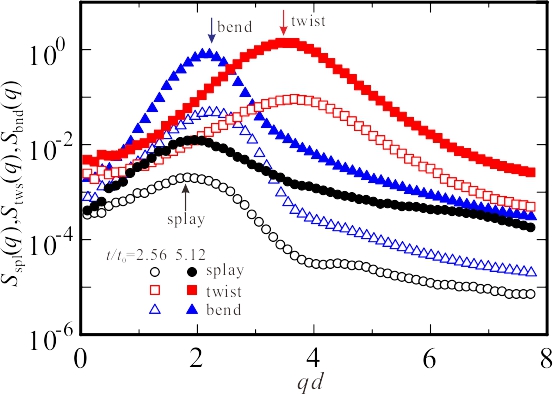} 
\end{center}
\caption{
\label{fig:spinodal}
(color online) 
Decomposed structure factors of the nematic order parameter 
in isotropic-nematic phase transitions at $t=2.56t_0$ 
(empty symbols) and $5.12t_0$ (full symbols). 
Black circles, red squares, and blue triangles represent the splay, 
Twist, and bend modes, respectively. 
Peak positions predicted by the linearized analysis are 
marked by the arrows. 
}
\end{figure}

Figure \ref{fig:spinodal} shows the decomposed structure 
factors of $Q_{ij}$ at $t=2.56t_0$ and $5.12t_0$. 
Their shapes are similar to those found in the spinodal decomposition of 
phase separations. 
Namely, each structure factor has a peak at an intermediate wave number 
and vanishes for $q \sim0$. 
Neglecting the higher order terms, 
we obtain the early-stage growth rates of the three modes in 
Eqs. (\ref{eq:S_spl})-(\ref{eq:S_bnd}), 
as denoted in Appendix C. 
The positions of the peaks, predicted by the linearized analyses, 
are marked by arrows in Fig. \ref{fig:spinodal}. 
The simulation results are consistent 
with the linearized theory and the splay mode develops more slowly than the other two modes. 

The linearized analysis indicates 
that the fluctuation of $\phi$ and the 
splay mode are coupled to each other. 
However, our simulated structure factor $\phi$ does not show 
an appreciable peak (data not shown). 
We consider that this is an artifact of our numerical simulation, 
because the complex spatial operators in the kinetic equations 
are difficult to deal with precisely. 
We need to improve the numerical scheme 
to study the spatial distribution of $\phi$ more quantitatively.

\subsection{Growth of polydomain for $L_{\rm r}=0$}

\begin{figure}[htbp]
\begin{center}
\includegraphics[angle=0,width=0.5\textwidth, bb= 0 0 393 709]{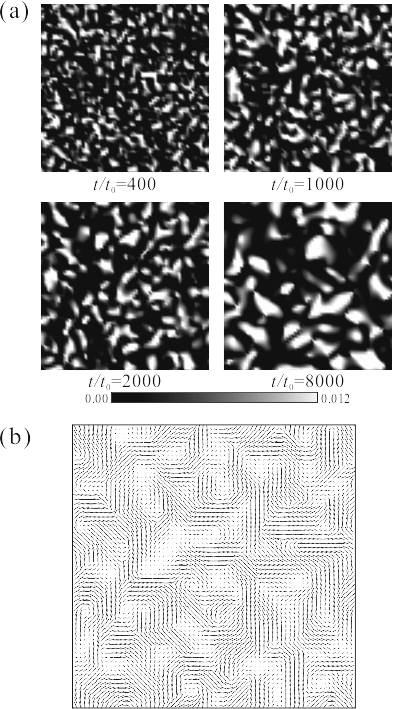}
\end{center}
\caption{
\label{fig:Schlieren}
(a) Snapshots of $Q_{xy}^2$ in an $xy$ plane ($z=0$) at 
$t/t_0=400,\,1000,\,2000$, and $8000$ in 
the isotropic-nematic transition without rotational diffusion. 
(b) The corresponding director field of the domain pattern 
in the same $xy$ plane at $t=8000t_0$. 
The lines represent the principle axes of the tensorial order parameters. 
}
\end{figure}

\begin{figure}[htbp]
\begin{center}
\includegraphics[angle=0,width=0.5\textwidth,bb= 0 0 1348 837]{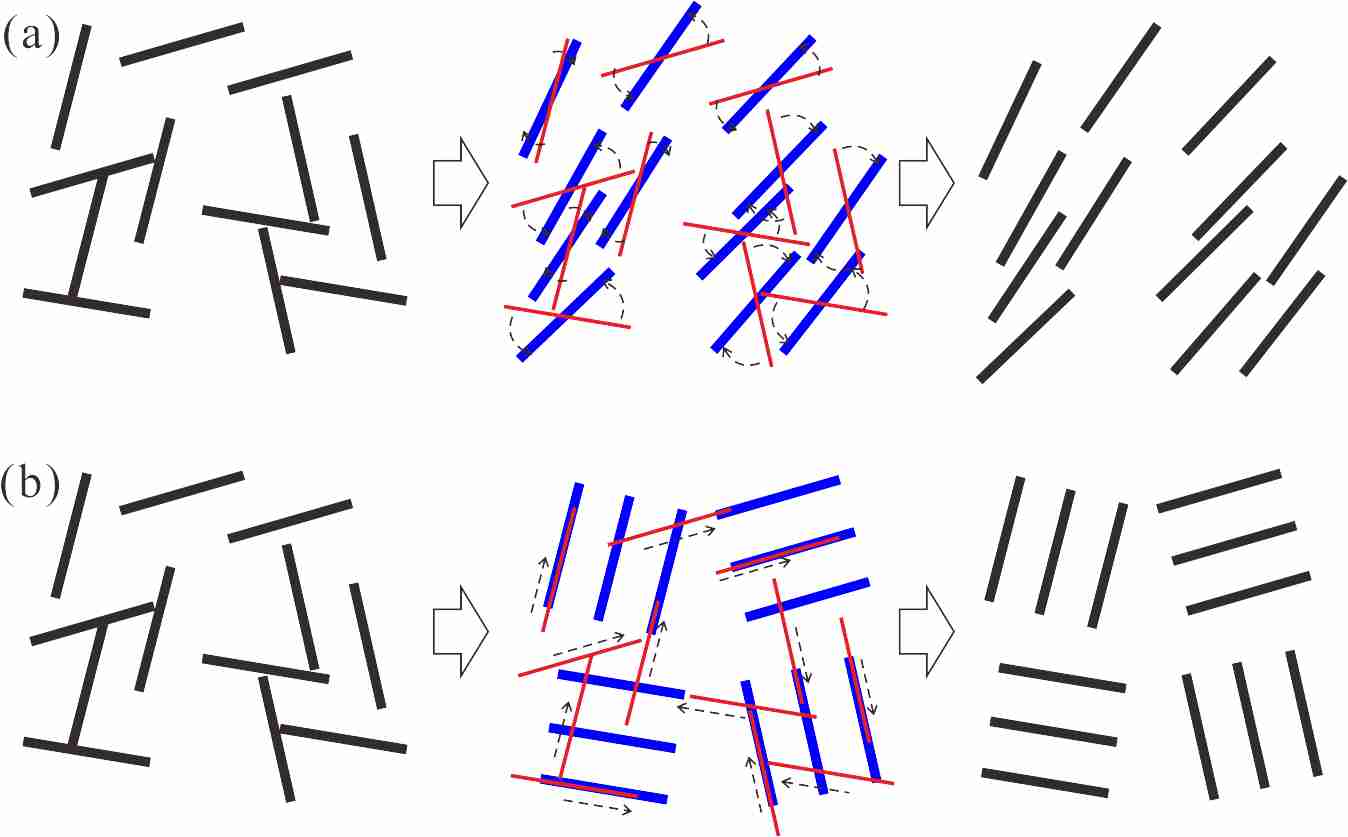}
\end{center}
\caption{
\label{fig:ordering}
(color online) 
Process of isotropic-nematic transition in a solution of rodlike polymers. 
(a) When the rotational motion is allowed, the polymers 
rotate to align with fixed positions as in LMWLCs. 
(b) In the absence of rotational motion, 
the rodlike molecules diffuse only along their rod direction to form 
small grains of the nematic phase. 
In the center column, red and blue segments represent 
initial (left) and ordered (right) configurations, respectively. 
Arrows of broken lines indicate the motions of the rodlike polymers. 
}
\end{figure}

Next, we study the temporal evolution of the polydomain pattern in 3D. 
We set $\alpha=16v_0$, $\Delta x=d$, and $\Delta t=0.05t_0$. 
The initial condition is $\bar{\phi}=0.5$, which is 
larger than $\phi_2$, and $\bar{Q}_{ij}=0$ with the random noise.
Figure \ref{fig:Schlieren}(a) shows the temporal 
evolution of the pattern of $Q_{xy}^2$ in an $xy$-plane ($z=0$). 
The corresponding director field at $t=8000t_0$ 
in the same $xy$-plane is shown in Fig. \ref{fig:Schlieren}(b). 
A polydomain pattern is formed 
and it coarsens in time. 
Figure \ref{fig:ordering} schematically 
explains how the isotropic-nematic transition 
takes place without molecular rotation. 
If the rotational motion is allowed, 
the rodlike polymers rotate to align 
with fixed positions as shown in Fig. \ref{fig:ordering}(a). 
On the other hand, when the rotational motion is severely suppressed,
the rodlike polymers that are initially nearly parallel to each 
other 
assemble to form a small grain via diffusion along their axes. 
The assembled grains form a mosaic pattern and 
many defects remain as shown in Fig. \ref{fig:ordering}(b). 
In the latter case, the 
nematic order parameter is conserved in the whole system. 

The defects in Fig. \ref{fig:Schlieren} are at the intersections of 
disclination lines in the $xy$-plane ($z=0$),
entangled in three dimensions. 
In 2D Schlieren textures, the number of bright brushes forming 
a defect core is given by $4|m|$, 
where $m$ is the topological strength of the defect. 
We observed that most defects have two brushes in Fig. \ref{fig:Schlieren}(a). 
As in other nematic states of LMWLCs, disclination lines of $m=\pm 1/2$ are 
formed more frequently than other types of defects.

Since the molecular rotation is severely suppressed in LCPs, 
its coarsening mechanism is very different from that in LMWLCs. 
Even after the early stage, the scalar nematic order parameter, 
which is given by $Q^0=(2 Q_{ij}Q_{ji}/3)^{1/2}$, 
remains inhomogeneous; 
usually, $Q^0$ is smaller than the equilibrium 
nematic order $Q^{\rm eq}$ near the defects.
Since the inhomogeneity of $Q^0$ affects the structure factor 
in the high $q$-range, we calculate the structure factor 
of a normalized order parameter, $\hat{Q}_{ij}=Q_{ij}/Q^0$, 
to determine the evolution of the polydomain pattern. 
This normalization corresponds to a binarization 
method for phase separation \cite{Shinozaki1993}. 

\begin{figure}[htbp]
\begin{center}
\includegraphics[angle=0,width=0.5\textwidth, bb= 0 0 599 835]{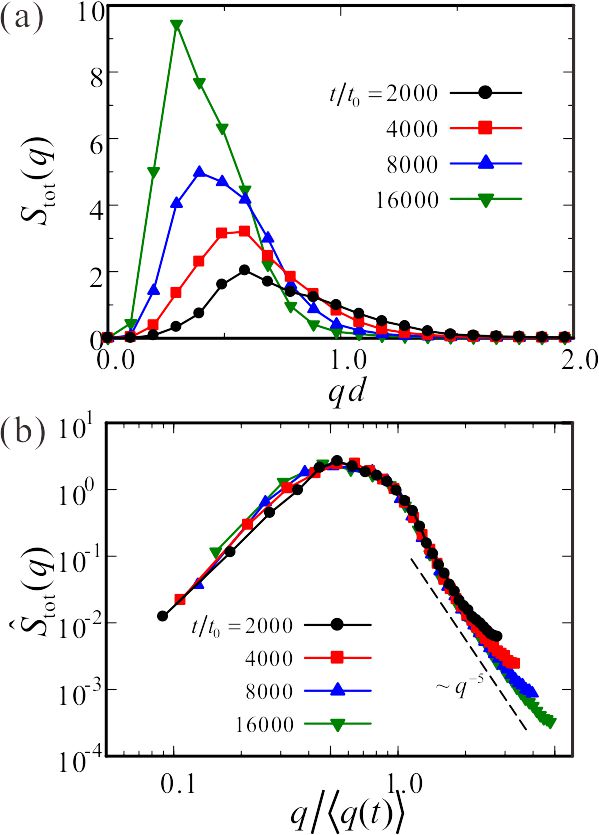}
\end{center}
\caption{
\label{fig:S(q)}
(color online) 
(a) Temporal changes in the total structure factor of the nematic 
order parameter in isotropic-nematic transitions without 
rotational diffusion. 
(b) Structure factor $\hat{S}_{\rm tot}(q)$ scaled by the characteristic wave number 
$\av{q(t)}$ of the domain pattern. 
$\hat{S}_{\rm tot}(q)$ shows the domain 
growth with self similarity and 
the Porod law $\hat{S}_{\rm tot}\sim q^{-5}$ is observed in the high 
$q$-regime. 
}
\end{figure}

In Fig. \ref{fig:S(q)}(a), we plot the temporal change of 
the total structure factor $S_{\rm tot}(q)=|\tilde{Q}_{ij}(\bi{q})|^2$, 
in which $\tilde{Q}_{ij}(\bi{q})$ refers to 
the Fourier transform of $\hat{Q}_{ij}(\bi{r})$. 
The structure factor is not decomposed into the three modes given 
by Eqs. (\ref{eq:S_spl})-(\ref{eq:S_bnd}). 
It is shown that the peak position of the structure factor shifts toward 
$q=0$ and the peak height develops with time. 
These features are similar to those in 
the late stage of phase separation 
\cite{Onukibook,Shinozaki1993}. 
It is known that the structure factor scaled by 
the characteristic wave number collapses into 
a master curve at isotropic phase separation. 
We replot the scaled structure factor, 
$\hat{S}_{\rm tot}(q)=\av{q(t)}^{3}S_{\rm tot}(q/\av{q(t)})$, 
in Fig. \ref{fig:S(q)}(b) and 
the characteristic wave number is defined as 
\bea
\av{q(t)}=\frac{\int d\bi{q} S(\bi{q},t)_{\rm tot}|\bi{q}|}
{\int d\bi{q}S_{\rm tot}(\bi{q},t)}.
\ena
Figure \ref{fig:S(q)}(b) shows 
that the dynamic scaling law holds fairly well in the 
isotropic-nematic transition of rodlike polymers. 

In the high wave number range ($q > \av{q(t)})$, 
the structure factor decays as $\tilde{S}_{\rm tot}(q)\sim q^{-5}$. 
In phase separation, a decay $S(q)\sim q^{-4}$, 
termed as Porod's law, which originates from 
the scattering of 2D interfaces in a 3D matrix, has been observed \cite{Onukibook}. 
 
The $q^{-4}$-tail was also reported in 
2D simulations of a nematic phase 
\cite{Zapotocky1995, Denniston2001}.
The $q^{-5}$-tail observed in LCPs is considerably 
different from these tails.
In similar systems, Bray studied phase transitions 
 described by a conserved $N$-vector order parameter \cite{Bray1990}, and 
showed that the structure factor exhibited a $q^{-(N+3)}$-tail in 3D. 
From his work, it follows that a $q^{-5}$-tail is obtained for a system of $N=2$ or $XY$ model, in which many line defects are formed. 
We consider that the $q^{-5}$-tail observed in LCPs 
represents scattering from the entangled one dimensional (1D) disclination lines 
in the 3D matrix.

\begin{figure} [htbp]
\begin{center}
\includegraphics[angle=0,width=0.5\textwidth,bb= 0 0 535 409]{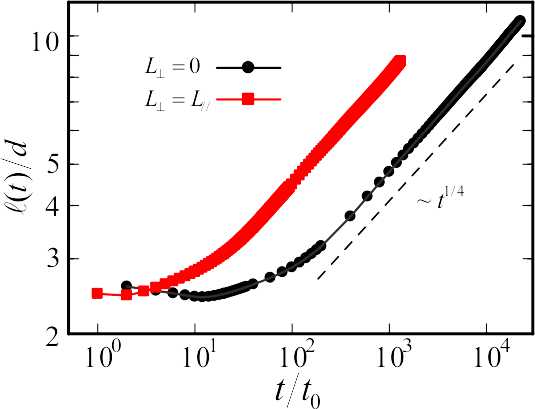}
\end{center}
\caption{
\label{fig:l-t}
(color online) 
Time evolutions of the characteristic domain length $\ell(t)$. 
Black and red symbols represent those for $L_\perp=0$ 
and $L_\perp=L_\parallel$, respectively. 
Rotational diffusion is set to $L_{\rm r}=0$. 
In both the cases, characteristic lengths grow with time 
according to $\ell(t)\sim t^{1/4}$. 
}
\end{figure}

The self-similarity in the scaled structure factors 
indicates that the polydomain growth is characterized by 
only one characteristic length scale. 
We plot the temporal change of the characteristic polydomain size, 
which is defined as $\ell(t)=2\pi/\av{q(t)}$ in Fig. \ref{fig:l-t}. 
After the early stage, the characteristic size develops with time 
as $\ell(t)\sim t^{\alpha}$ with $\alpha=1/4$. 
This exponent is smaller than those 
in phase separation ($\alpha=1/3$) and 
isotropic-nematic transition of LMWLCs ($\alpha=1/2$) 
\cite{Toyoki1994,Fukuda1998,Denniston2001}. 
Interestingly, 
the exponent is the same as that of a system 
described by a conserved $XY$ model \cite{Bray1990,Siegert1993}. 
This coincidence and the same Porod's tail value
imply profound similarities between LCPs and the conserved 
$XY$ model. 
Therefore, the analysis of the 
conserved $XY$ model might be helpful to 
understand the coarsening mechanism in LCPs. 
However, there is also an important difference between the two; 
while most defects have $m=\pm 1/2$ topological strengths 
in the nematic state of rodlike polymers, 
the topological strengths in the $XY$ model are $m=\pm 1$. 
Further studies are needed to clarify the similarities and 
differences between them. 

We have also studied the effect of the perpendicular diffusion 
on the polydomain growth. 
We set $L_\perp=L_\parallel$ and $L_{\rm r}=0$, 
where the off-diagonal terms of the kinetic equations 
(\ref{eq:develop_phi2}) and (\ref{eq:develop_Q2}) vanish. 
Since the rotational diffusion is not included, 
the tensorial order parameter $Q_{ij}$ is still conserved. 
The numerical simulations show that the dynamic scaling law 
holds and the growth exponent is also given by $\alpha=1/4$ 
as shown in Fig. \ref{fig:l-t} (red squares). 
It is indicated that this growth exponent is not 
characteristic of the parallel diffusion, 
but stems from the nature of the preserved order parameter.
The characteristic length for $L_\perp=L_\parallel$ grows faster with time 
than for $L_\perp=0$ by a factor of approximately 1.6.

\subsection{Growth of polydomain for $L_{\rm r}> 0$}

\begin{figure} [htbp]
\begin{center}
\includegraphics[angle=0,width=0.5\textwidth,bb= 0 0 531 408]{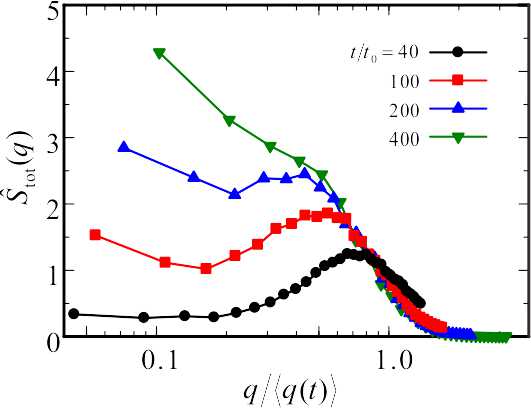}
\end{center}
\caption{
\label{fig:S(q)2} 
(color online) 
Time development of the scaled structure factor, $\hat{S}_{\rm tot}(q)$, 
in the isotropic-nematic transition with 
rotational diffusion, $L_{\rm r}d^2=0.0625 L_\parallel$. 
Gradual evolution in the low $q$-regime is observed after the crossover time $t_{\rm cr}\cong 100t_0$. 
}
\end{figure}

Solutions of LCPs have a very small but finite 
rotational diffusion coefficient $D_{\rm r}$. 
Thus, we expect that the polydomain growth will be affected 
by the rotational diffusion in the late stage of 
phase ordering. 
Figure \ref{fig:S(q)2} shows the temporal change of 
the scaled structure factor, $\hat{S}_{\rm tot}(q)$. 
Here, we set $L_\perp=0$ and $L_{\rm r}d^2/L_{\parallel}=0.0625$ and 
the other parameters are the same as those 
for $L_{\rm r}=L_\perp=0$. 
In the early stage, the structure factor has 
the same features as those in Fig. \ref{fig:S(q)}, 
namely, $\hat{S}_{\rm tot}(q)$ is very small at $q\sim 0$ 
and it has a peak at an intermediate wave number. 
With time, the structure factor 
in the lower $q$-range develops and in the late stage, $\hat{S}_{\rm tot}(q)$ has the 
Ornstein-Zernike form $S(q)\sim (1+q^2\ell^2)^{-1}$, 
which is also observed in LMWLCs. 
The dynamic scaling law does not hold during the 
whole phase transition process and the 
growth of $\hat{S}_{\rm tot}(q)$ in the lower $q$-range 
is attributed to the rotational motion 
of rodlike polymers.

\begin{figure}[htbp]
\begin{center}
\includegraphics[angle=0,width=0.5\textwidth, bb=0 0 534 413]{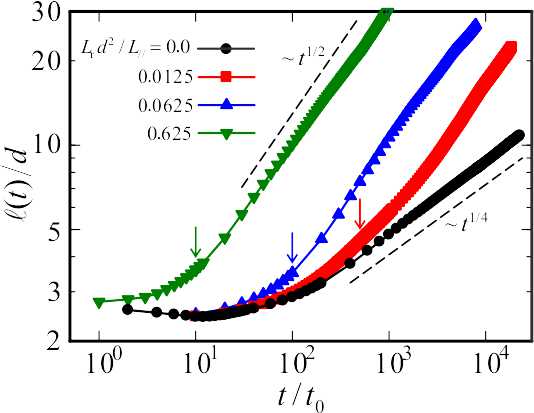}
\end{center}
\caption{\label{fig:l-t2} 
(color online) 
Temporal changes in the characteristic domain lengths in the 
isotropic-nematic transitions in LCPs. 
Rotational diffusion constant is varied as 
$L_{\rm r}d^2/L_\parallel=0,\,0.0125,\,0.0625$, and $0.625$. 
The occurrence of crossovers is indicated by arrows , 
and their positions are given by $t_{\rm r}=6.25/D_{\rm r}$. 
}
\end{figure}

In Fig. \ref{fig:l-t2}, we show the time evolution of 
$\ell$ for a number of $L_{\rm r}$'s. 
Although the physical meaning of $\ell$, 
especially in the crossover period (see below), is not clear, 
it is still a useful measure for the pattern growth. 
In the early stage, it does not change with time. 
This steady length corresponds to the spinodal 
decomposition-like growth of the nematic order parameter. 
After the early stage, the domain length evolves obeying $\ell\sim t^{1/4}$ 
as in the case of $L_{\rm r}=0$. 
Figure \ref{fig:l-t2} indicates that 
the growth exponent changes from $\alpha=1/4$ to $\alpha=1/2$, 
which is the same as in the case of phase transition of LMWLCs, 
during a crossover period. 
Fig. \ref{fig:l-t2} also indicates that the crossover 
depends on the rotational diffusion constant. 
As $L_{\rm r}$ increases, the crossover is observed at earlier times. 

Each rodlike polymer moves along a tube surrounded by other tubes. 
The tubes are not necessarily straight and they disappear in a certain period of time, 
New tubes are continually created 
as the surrounding molecules fluctuate. 
As a result, the polymer gradually loses 
its original orientation \cite{Doi1984}. 
We estimate that the crossover time $t_{\rm cr}$ is 
of the order of the characteristic rotational time 
$D_{\rm r}^{-1}$. 
After the crossover, the orientational order parameter is 
no longer conserved. 
In Fig. \ref{fig:l-t2}, arrows 
mark the corresponding crossover times and we assume $t_{\rm cr}=6.25t_0L_\parallel /(L_{\rm r}d^2)$. 
Although it is difficult to determine exactly 
when the exponent changes from $\alpha=1/4$ to 
$\alpha=1/2$, the values marked by the arrows appear to 
be consistent with this interpretation. 

\begin{figure}[htbp]
\begin{center}
\includegraphics[angle=0,width=0.5\textwidth, bb=0 0 571 648]{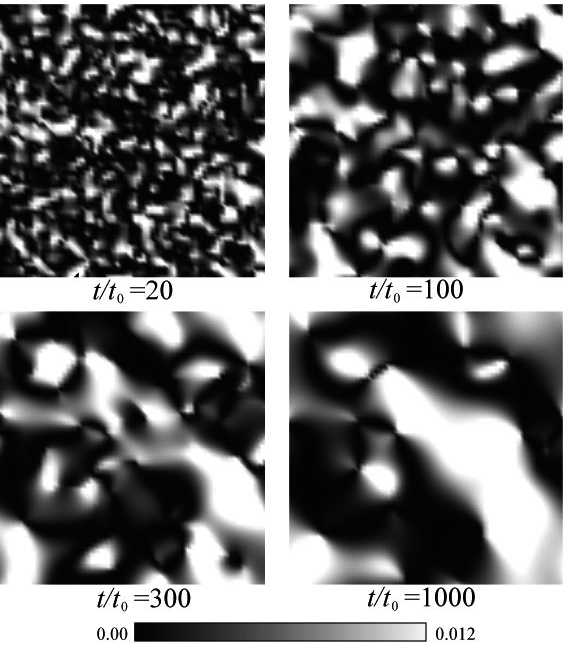}
\end{center}
\caption{
\label{fig:Schlieren2}
Snapshots of $Q_{xy}^2$ in the $xy$ plane after the crossover. 
Rotational diffusion constant is $L_{\rm r}d^2=0.625L_{\perp}$. 
}
\end{figure}

In Fig. \ref{fig:Schlieren2}, we show the time evolution 
of the Schlieren pattern $Q_{xy}^2$ in an $xy$-plane 
for $L_{\rm r}d^2/L_\parallel=0.625$ and the crossover time is estimated as $t_{\rm cr} \cong 10t_0$. 
In an LCP solution with a finite $L_{\rm r}$, 
the director field rotates slowly, but freely, 
to adjust to the surrounding molecules.

\subsection{Defect motion}

We study the finite-size effects on the stability of defects, mediated by 
the elastic field of the nematic phase 
for a defect pair of anti-signed topological charges. As there is an attractive interaction between the charges, 
the defects approach and annihilate each other. 
It was reported that 
the separation of $R$ between the defects
decreases to zero with time as 
$R\sim (t_{\rm a}-t)^{\beta}$ with $\beta=1/2$, 
where $t_{\rm a}$ is the annihilation time 
\cite{Zapotocky1995,Toth2002,Svensek2002}. 
After annihilation, 
the director field relaxes to a homogeneous state 
in order to release the elastic energy in LMWLCs. 
However, 
this argument is not applicable to LCPs, 
since the kinetic mechanism is different. 

In the simulations, $\alpha=16v_0$, 
$\bar{\phi}=0.5$, $\Delta x=0.2 d$, and $\Delta t=0.0002t_0$, and 
we set $L_\perp=0$ and $L_{\rm r}=0$ to consider 
only parallel diffusions, i.e., those along the polymer molecular axis. We place a pair of anti-signed defects ($m=\pm 1/2$) 
in a small square box and carry out 2D simulations
to mimic two parallel disclination lines 
with topological strengths of $m=\pm 1/2$. 

Initially, the spatial distribution of $Q_{ij}$ is set to 
$Q_{ij}=3Q^{\rm eq}\left(u_i u_j-\delta_{ij}/3\right)/2$ 
with 
\bea
\bi{u}(\bi{r})=(\cos\{(\theta_+-\theta_-)/2\},
\sin\{(\theta_+-\theta_-)/2\},0).
\ena 
Here, $\theta_+(\bi{r})$ is the angle between 
$(\bi{r}-\bi{r}_+)$ and $(\bi{r}_--\bi{r}_+)$, 
and {\it vice versa}. 
$\bi{r}_+$ and $\bi{r}_-$ are the defect positions 
of $m=1/2$ and $m=-1/2$, respectively \cite{Toth2002}. 
At $t=0$, $(\bi{r}_++\bi{r}_-)/2$ is at 
the center of the box and $\bi{r}_+-\bi{r}_-$ is along the 
$x$-axis. 
This configuration is not the equilibrium 
structure of the defect positions, because the 
spatial variations of the scalar 
nematic order parameter and the non-linear terms are neglected. 
The absence of these terms do not affect our results, 
since the structure relaxes quickly before defect motions are excited. 
In order to avoid numerical artifacts near the boundary wall, 
we employ the periodic boundary condition; this
may be inappropriate to study the
realistic confinement effects on the defect motion; however, as 
there are only two defects in the system, 
the simulations give 
valuable insights into the stability of the defects.

\begin{figure} [htbp]
\begin{center}
\includegraphics[angle=0,width=.7\textwidth, bb=0 0 1539 1075]{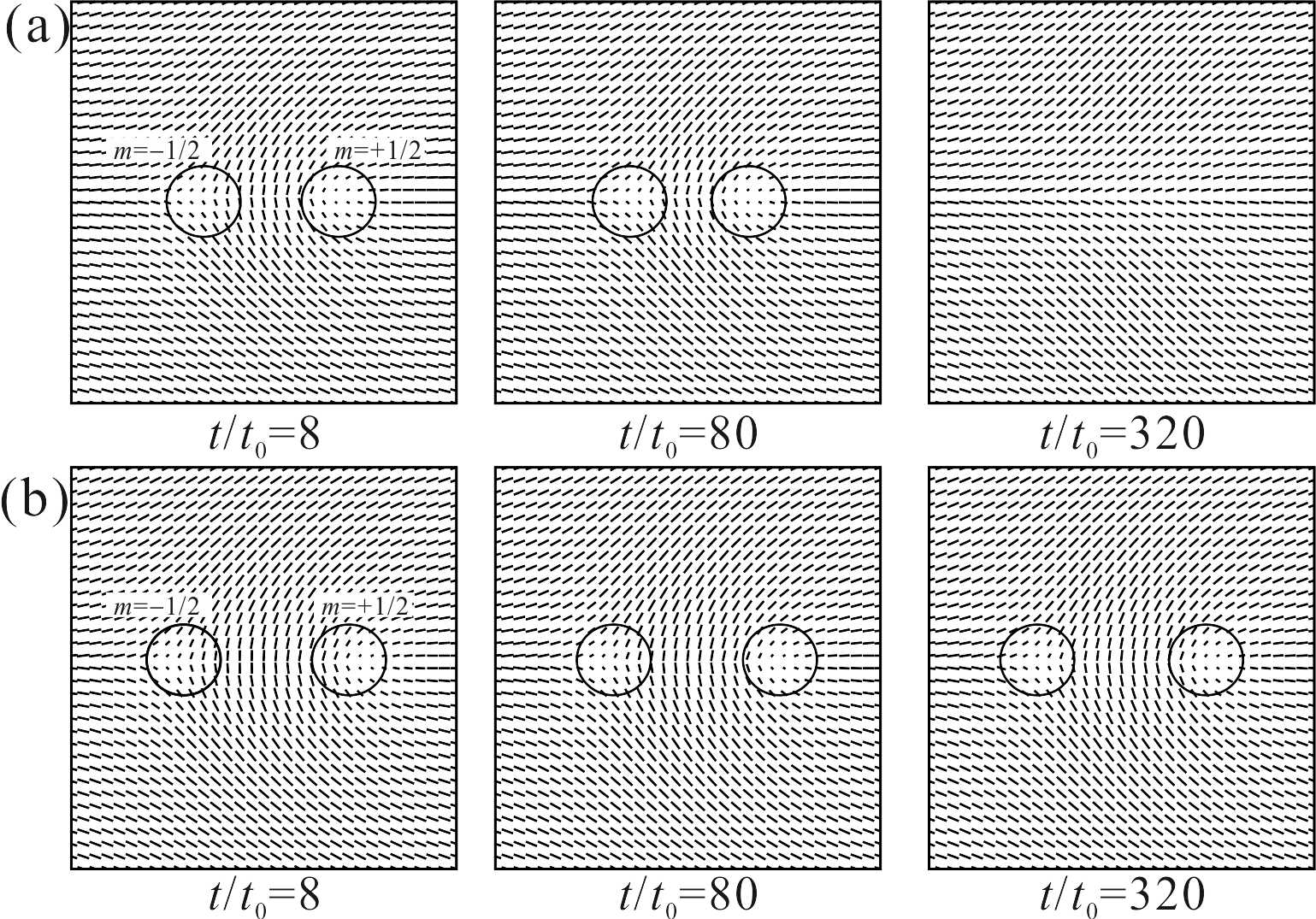}
\end{center}
\caption{
\label{fig:defect} 
Director fields around a pair of anti-signed defects at 
$t/t_0=8$, $80$, and $320$. 
The initial separation between the defects is (a) $R(0)=12\Delta x$ 
and (b) $R(0)=16\Delta x$. 
The square box size is $H=48\Delta x$ and rotational diffusion is not allowed. 
Owing to the conservation of the order parameter, the director 
field remains deformed even after the defects are annihilated. 
At large separations, the defect structure is stabilized.
}
\end{figure}

Figure \ref{fig:defect} 
shows time evolution of 
the director fields in $Q_{ij}$. 
In Fig. \ref{fig:defect}(a), 
the initial defect separation is $R(0)=|\bi{r}_+-\bi{r}_-|=12\Delta x$ 
and the box is $H^2$, where $H=48\Delta x$. 
The two defects approach and annihilate each other at $t\cong 153.3t_0$. 
Contrary to the cases in LMWLCs,
the director field remains distorted 
even after a long annealing time. 
This is because the $y$-component of the director field 
is preserved. 
In other words, 
the rodlike polymers, which are initially oriented along the $y$-axis,
remain permanently aligned with the axis if $L_{\rm r}=0$.

The free energy of the final distorted state 
in Fig. \ref{fig:defect}(a) 
is lower than that of a uniform nematic state with 
the same average order parameter $\bar{Q}_{ij}$. 
Here, it is important that $\bar{Q}_{ij}$ 
differs from the equilibrium value $Q^{\rm eq}_{ij}$. 
The final distorted state is determined by the 
balance between the local and non-local terms 
in the free-energy functional [Eqs. (\ref{eq:F0_2}) 
and (\ref{eq:free_ene2})]. 
For LMWLCs, the director field in the equilibrium state can 
optimize both parts of the free energy, 
such that $\bar{Q}_{ij}=Q^{\rm eq}_{ij}$. 
On the other hand, as the lowering of the local part has to 
induce the deformation of the director field in LCPs, 
it is reasonable to assume that the elastic distortion remains 
even in the final state.

\begin{figure} [htbp]
\begin{center}
\includegraphics[angle=0,width=0.5\textwidth,bb= 0 0 533 401]{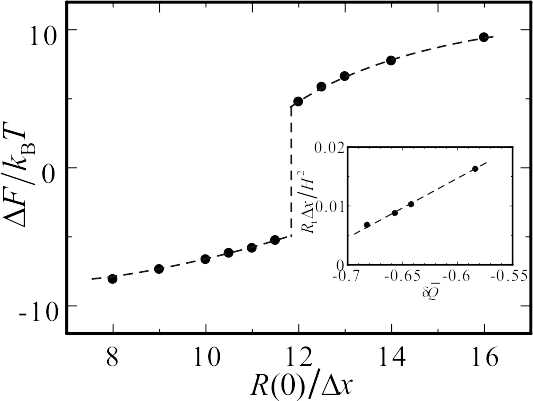}
\end{center}
\caption{
\label{fig:defect-ene} 
Plot of the free-energy difference 
at $t=800t_0$ with respect to 
the initial separation of the defect pair. 
The box size is fixed at $H=32\Delta x$ and the 
jump around $R(0)\cong 11.5\Delta x$ represents the threshold 
$R_{\rm t}$, above which defects remain even after a long annealing time. 
Inset: reduced threshold $R_{\rm t}\Delta x/H^2$ vs. 
 $\delta \bar{Q}$. 
}
\end{figure}

Figure \ref{fig:defect-ene} plots the free energy at $t=800t_{0}$ 
calculated by Eqs. (\ref{eq:F0_2}) and 
(\ref{eq:free_ene2}) versus the initial 
separation in a square box of $H=32\Delta x$. 
With increasing $R(0)$, the free energy is 
increased for $R(0)\lesssim 11\Delta x$. 
This increase enhances 
the elastic distortion without the formation of defects, 
as observed in Fig. \ref{fig:defect}. 
A kink in the free energy is observed around $R\cong 11.5\Delta x$. 
In Fig. \ref{fig:defect-ene}, 
the free-energy difference from a 
reference 
$\{\mathcal{F}(11\Delta x)+\mathcal{F}(12\Delta x)\}/2$ is 
plotted. 
This kink represents the critical separation $R_{\rm t}$, 
above which 
the defects do not annihilate each other and remain even after 
a long annealing time ($t=800t_0$). 
The snapshots of the stabilized director field 
for $R(0)=16\Delta x$ in the box of $H=48\Delta x$, 
are shown in Fig. \ref{fig:defect}(b). 
These stable defects also stem from the conservation of the order parameter. 
As the initial separation increases, the amount of the rodlike polymers 
oriented along the $y$-axis increases, 
so that the elastic field is appreciably distorted. 
Above the critical separation, 
the formation of local singular points (defects) 
are preferable to gradual distortion without defects. 

\begin{figure}[htbp]
\begin{center}
\includegraphics[angle=0,width=0.5\textwidth, bb= 0 0 577 812]{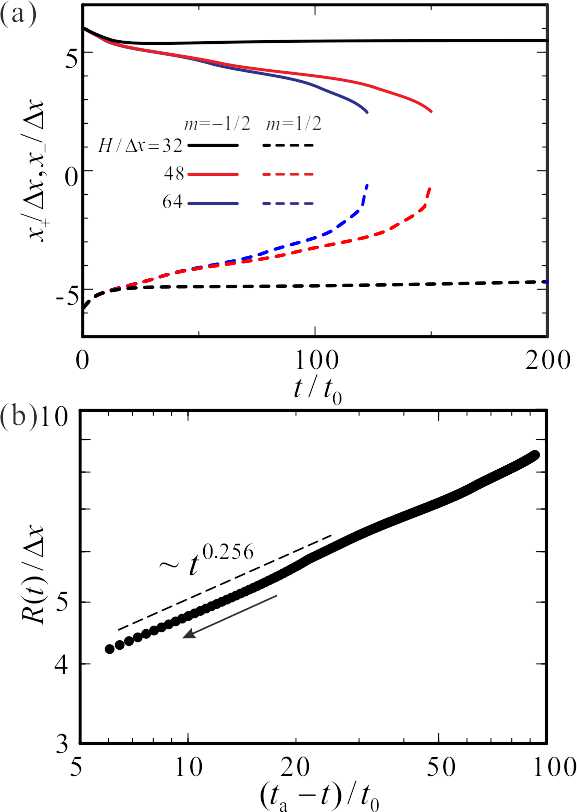}
\end{center}
\caption{
\label{fig:defect-sep} 
(color online) 
(a) Time evolutions of the positions to the right ($m=1/2$) and 
left ($m=-1/2$). 
Initial separation is $R(0)=12\Delta x$ and the box size 
is varied as $H/\Delta x=32,\,48$, and $64$. 
When $H=32\Delta x$, the defect positions are almost fixed 
after a small shift in the early stage. 
(b) Separation of the defect pair as a function of $t_{\rm a}-t$, 
where $t_{\rm a}$ is the annihilation time. 
Initial separation and the box size are $R(0)=12\Delta x$ 
and $H=48\Delta x$, respectively.
The curve is fitted with the function $(t_{\rm a}-t)^\beta$, 
where $t_{\rm a}=153.3t_0$ and $\beta=0.256$. 
}
\end{figure}

The critical separation depends on the system size. 
Figure \ref{fig:defect-sep}(a) shows 
the defect positions as a function of time. 
We fix the initial separation to $R=12\Delta x$ 
and vary the system size by $H/\Delta x =32$, $48$, and $64$. 
In the largest system, the defects approach faster and 
as $H$ decreases, the defect motion becomes slower and 
the resultant annihilation time is retarded. 
For $H=32\Delta x$, the defects initially experience a small shift at early 
times ($t\lesssim 10t_0$) and then hardly move.
This dependence on the system size is unique to LCPs, and is not observed in LMWLCs. 
When the system size is large, there is a lot of room for
 the incompatible rodlike polymers to diffuse. 
In the inset of Fig.\ref{fig:defect-ene}, 
we show the dependence of the threshold $R_{\rm t}$, 
on the average order parameter difference, 
$\delta \bar{Q}=\int d\bi{r}(Q_{yy}-Q_{xx})/H^2$. 
In the initial configuration, the rodlike polymers along the $y$-axis 
are localized in between the two defects; therefore, 
the total amount of polymers is expected to be 
proportional to the defect separation, 
{\it i.e.,}, $\int d\bi{r}(Q_{yy}-Q_{xx})\propto R$. 
When $|\delta \bar{Q}|$ is smaller than a critical value, 
the solution cannot relax to a homogeneous nematic phase 
without defects and the simple scaling relation for the 
critical defect separation is $R_{\rm t}/H^2\propto \delta\bar{Q}$.

This size dependence is similar to that in 
systems described by a single scalar order parameter. 
Here, we consider a system whose free energy 
has two minima below its critical point. 
We assume that a droplet of one phase is placed 
in a matrix of the other phase and 
if the order parameter is not conserved, 
the droplet will be adsorbed in the matrix phase, as in magnetism. 
When the order parameter is conserved, as in phase 
separation, the droplet can stably exist 
in a confined system. 
In its steady state, the radius of the droplet is determined 
by the average volume fraction of the components, and the 
 concentration of the matrix phase is slightly 
supersaturated compared to the equilibrium concentration. 
This supersaturation is related to the interface tension 
given by the Gibbs-Duhem relationship \cite{Onukibook}. 
As the volume of the confined box increases with a fixed droplet radius, 
the droplet evaporates and the system becomes homogeneous. 
This is because supersaturation decreases with increasing 
box size and the resultant critical droplet size is increased. 
In LMWLCs, the supersaturation relaxes locally and quickly to the 
equilibrium state. 
On the other hand, in LCPs, the system is ``supersaturated" 
from the equilibrium state and the large supersaturation leads to 
the deformation of the director field.

In Fig. \ref{fig:defect-sep}(b), we replot the defect separation 
as a function of reduced time, $t_{\rm a}-t$.
The initial separation is $R=12\Delta x$ in a square box 
of $H=48\Delta x$, and 
the defects annihilate each other at $t=t_{\rm a}(\cong 153.3t_0)$. 
We estimate the exponent $\beta\cong 0.256$ 
in $R\sim (t_{\rm a}-t)^\beta$ by fitting the curve. 
This is considerably smaller than $\beta=0.5$ 
for LMWLCs and close to the growth exponent $\alpha$. 
Note that the annihilation exponent $\beta$ is 
also of the same value as the growth exponent $\alpha$ in LMWLCs; 
however, the mechanism is not clearly understood and we 
have not concluded whether this exponent is universal for LCPs. 
Interestingly, Fig. \ref{fig:defect-sep}(a) also suggests that 
the defect motion before the annihilation becomes asymmetric; 
the defect for $m=1/2$ moves faster than the defect for $m=-1/2$. 
Asymmetric motions of defects are also observed 
in LMWLCs with hydrodynamic interactions \cite{Toth2002}; 
however, these interactions are absent in our model.

\section{Conclusion}

We studied the isotropic-nematic transition in liquid crystalline 
polymers by integrating the time-dependent Ginzburg-Landau equations 
for the compositional order parameter $\phi$ and 
the orientational order parameter $Q_{ij}$. 
The kinetic coefficients are evaluated using the 
Fokker-Planck equation \cite{Shimada1988c}. 
This approach ensures that
the rodlike polymers diffuse only 
in the direction parallel to their molecular axis. 

Even if rotational motion is not allowed, 
the polymers being nearly parallel to each other 
assemble to form a nematic grain in the 
early stage of the phase transition. 
Since the director field is randomly oriented 
before the quenching of the isotopic phase, a polydomain structure 
is formed and many defects remain at the grain boundaries. 
The polydomain growth then exhibits self-similarity and 
the growth exponent is $\alpha=1/4$, which is 
considerably smaller than that for nematic liquid crystals of 
low molecular weight molecules. 
This small exponent is similar to those found 
in systems with conserved vector order parameters \cite{Bray1990}. 
Here, the structure factor of the nematic order parameter 
has a peak at an intermediate wave number 
and nearly vanishes at $q\sim 0$. 
Hence, the orientational order parameter is preserved, 
contrary to that in LMWLCs. 

We have also shown that small but finite rotational diffusions 
can dominate the dynamics after a crossover time $t_{\rm cr}$. 
After the crossover, the growth exponent changes to $\alpha=1/2$, 
which is same as that for LMWLCs, and
the polydomain pattern and the structure factor 
are also similar to those of LMWLCs. 
We estimated the crossover time as $t_{\rm cr}\cong 1/D_{\rm r}$, 
which enables rodlike polymers to rotate and 
the system behaves as a normal nematic liquid crystal, as in LMWLCs. 
This estimation qualitatively explains our numerical results.

We have also shown that defect motion is strongly 
influenced by the conservation of the nematic order parameter. 
In LCPs, the defects can be stabilized in a confined system 
and the stability depends on the box size. 
When the stability is lowered by increasing the box size, 
a pair of anti-symmetric defects annihilate each other. 
The director field is distorted 
after the annihilation of the defect pair and the
 defect annihilation obeys the power law, 
{\it i.e.,} $R\sim (t_{\rm a}-t)^\beta$ with $\beta\cong 0.25$. 

There are many experimental studies on the phase transition of LCPs. 
However, most have been analyzed 
with the same kinetic equation used for LMWLCs. 
Before the crossover is reached, 
the molecule diffuses by a crossover length 
$\ell_{\rm cr}=(D_\parallel/D_{\rm r})^{1/2}$. 
For a semi-dilute solution of rodlike polymers, 
the crossover length $\ell_{\rm cr}$ 
is comparable to the molecular length $d$. 
In order to examine the crossover in the phase transitions in LCPs, 
one must carefully probe the structure factor around 
$q\sim 2\pi/d$ at very early times. 
In a concentrated solution, 
the rotational diffusion constant is approximated 
by $D_{\rm r}=\beta D_{\rm r}^0(\bar{\phi} d^2/w^2)^{-2}$, 
where $\beta$ is the numerical factor \cite{Doi1984,Teraoka1985}. 
Hence, the crossover length is increased with the average volume 
fraction $\bar{\phi}$ as $\ell_{\rm cr}\sim \bar{\phi}d^3/w^2$. 
In a melt of LCPs or a highly concentrated solution of rodlike polymers, 
the crossover might be experimentally accessible, and 
we hope that our numerical study will stimulate detailed experimental 
observations in these systems in the near future.

\begin{acknowledgments}

The authors thank A. Onuki and C. P. Royall for their helpful 
discussions and critical readings of the manuscript. 
This work was supported by a grant-in-aid from the Ministry of Education, 
Culture, Sports, Science and Technology (MEXT) of Japan. 
The computational work was carried out using the facilities at 
the Supercomputer Center, Institute for Solid State Physics, 
University of Tokyo (Japan).

\end{acknowledgments}

\appendix

\section{Isotropic approximation}

When we derive the free-energy functional (Eq. (\ref{eq:free_ene2})) 
and kinetic equations [Eqs.(\ref{eq:develop_phi2}) and (\ref{eq:develop_Q2})], 
we employ the isotropic average approximation \cite{Shimada1988c}. 
The isotropic average of $A(\bi{u})$ is defined by 
\bea
\av{A}_{\bi{u}}=\frac{1}{4\pi}\int d\bi{u} A(\bi{u}).
\ena
Then, the following formulas are obtained: 
\bea
&&\av{u_i}_{\bi{u}}=\av{u_iu_ju_k}_{\bi{u}}=0,\\
&&\av{u_iu_j}_{\bi{u}}=\frac{1}{3}\delta_{ij},\\
&&\av{u_iu_ju_ku_l}_{\bi{u}}=\frac{1}{15}(\delta_{ij}\delta_{kl}
+\delta_{ik}\delta_{jl}+\delta_{il}\delta_{jk}).
\ena
Furthermore, we obtained 
\bea
&&\av{|\bi{u}\times\bi{u}'|}_{\bi{u}'}=\frac{\pi}{4},\\
&&\av{|\bi{u}\times\bi{u}'| u_i'u_j'}_{\bi{u}'}=
-\frac{\pi}{32}(u_iu_j-3\delta_{ij}). 
\ena
In deriving Eq. (\ref{eq:free_ene2}), we need higher-order moments 
of $\bi{u}$ such as $\av{|\bi{u}\times\bi{u}'|u_i'u_j'u_k'u_l'}_{\bi{u}}$, 
which are given in Ref \cite{Shimada1988c}.

\section{Excluded volume interaction between rodlike polymers}

In Eq. (\ref{eq:interaction}), $W(\bi{r},\bi{u};\bi{r}',\bi{u}')$ 
represents the excluded volume interaction potential between rodlike 
polymers $(\bi{r},\bi{u})$ and $(\bi{r}',\bi{u}')$.
In Fourier $\bi{k}$-space, it is expressed by \cite{Straley1973,Shimada1988b} 
\bea
\tilde{W}(\bi{k},\bi{u},\bi{u}')&=&\int d\bi{r}W(\bi{r},\bi{u},\bi{0},\bi{u}')
e^{i \bi{k}\cdot\bi{r}}\nonumber\\
&=&2\alpha \left|\bi{u}\times \bi{u}'\right|
\frac{\sin (\bi{K}\cdot\bi{r})}{\bi{K}\cdot\bi{u}}
\frac{\sin (\bi{K}\cdot\bi{u}')}{\bi{K}\cdot\bi{u}'}, 
\ena
where $\alpha$ is the interaction parameter that is estimated as $wd^2$. 
$\bi{K}$ is defined as $\bi{k}d/2$, and
$d$ and $w$ are the length and width of the rodlike molecules, respectively. 
We expanded it up to the order of $k^4$ using the following expansion: 
\bea
\frac{\sin(\bi{K}\cdot\bi{u})}{\bi{K}\cdot\bi{u}}=
1-\frac{(\bi{K}\cdot\bi{u})^2}{6}+\frac{(\bi{K}\cdot\bi{u})^4}{120}+\cdots. 
\ena
This approximation is allowed for $k\le 1/d$.

\section{Linear analysis}

Here, we derive the growth rates of the order parameters in the 
early stage of the isotropic-nematic transition. 
We assume $L_\perp=L_{\rm r}d^2=0$ in the high concentration limit. 
The deviations of the order parameters 
from the initial conditions 
are so small that 
the chemical potential and molecular field are linearized as 
\bea
\mu &\cong& \left(\frac{1}{\bar{\phi}}+\chi+C_0\bi{q}^2+C_1\bi{q}^4\right)
\delta \phi+K_0q_iq_j \delta Q_{ij},\\
H_{ij}&\cong& \left(\frac{a_0}{\bar{\phi}}-a_1\right)
\delta Q_{ij}+K_0q_iq_j\delta \phi
+K_1\bi{q}^2\delta Q_{ij}+\frac{1}{2}K_2
\left(q_iq_k \delta Q_{kj}+q_jq_k\delta Q_{ki}\right), 
\ena
where $\delta\phi=\phi-\bar{\phi}$ and $\delta Q_{ij}=Q_{ij}$. 

As noted in the main text, 
the orientational order parameter can be decomposed into 
the three modes splay, twist, and bend. 
According to Shimada {\it et al.}, 
the kinetic equations are also linearized as 
\bea
\pp{}{t}\delta \tilde{\phi}&=&-\Gamma_{\phi,\phi}(q)\delta
\tilde{\phi}
-\Gamma_{\phi,{\rm s}}(q)\delta \tilde{Q}_{\rm spl}\\
\pp{}{t}\delta \tilde{Q}_{\rm spl}&=&-
\Gamma_{{\rm s},\phi}(q)\delta\tilde{\phi}
-\Gamma_{\rm s,s}(q)\delta \tilde{Q}_{\rm spl}\\
\pp{}{t}\delta \tilde{Q}_{\rm tws}&=&-\Gamma_{\rm t,t}(q)
\delta \tilde{Q}_{\rm tws},\\
\pp{}{t}\delta \tilde{Q}_{\rm bnd}&=&-\Gamma_{\rm b,b}(q)
\delta \tilde{Q}_{\rm bnd}, 
\ena
where $\delta\tilde{Q}_{\rm spl}$, 
$\delta\tilde{Q}_{\rm tws}$, and 
$\delta\tilde{Q}_{\rm bnd}$ are 
the decomposed splay, twist, and bend modes of the tensorial 
order parameter $Q_{ij}$ in Fourier space, respectively. 
The decomposed structure factors in Sec. \ref{sec:spinodal} 
are obtained from them. 
For example, $\delta \tilde{Q}_{\rm spl}$ 
stands for $a_i\tilde{Q}_{ij}a_j$ in Eq. (\ref{eq:S_spl}). 
The coefficients are given by 
\bea
\Gamma_{\phi,\phi}(q)&=&\frac{1}{3}L_\parallel q^2
\left\{A+\left(C_0+\frac{2}{5}K_0\right)q^2+C_1q^4\right\},\\
\Gamma_{\phi,{\rm s}}(q)&=&\frac{2}{15}L_\parallel q^2\left\{
B+\left(\frac{5}{2}K_0+K_1+K_2\right)q^2\right\},\\
\Gamma_{{\rm s},\phi}(q)&=&\frac{4}{45}L_\parallel q^2\left\{
A+\left(C_0+\frac{11}{14}K_0\right)q^2+C_1q^4\right\},\\
\Gamma_{\rm s,s}(q)&=&\frac{22}{415}L_\parallel q^2\left\{
B+\left(\frac{14}{11}K_0+K_1+K_2\right)q^2\right\},\\
\Gamma_{\rm t,t}(q)&=&\frac{2}{35}L_\parallel q^2
\left\{B+\left(K_0+\frac{1}{2}K_1\right)q^2\right\}\\
\Gamma_{\rm b,b}(q)&=&\frac{2}{105}L_\parallel q^2(B+K_1q^2), 
\ena
where $A=\chi+1/\bar{\phi}$ and $B=a_0/\bar{\phi}-a_1$. 
Although $\Gamma_{\phi,\phi}$ and $\Gamma_{{\rm s},\phi}$ contain 
a negative $C_0$, both the growth rates 
increase monotonically with increasing $q$; therefore, 
 a modulated pattern due to negative $C_0$ will not appear. 

In the early stage, the concentration field and splay mode are 
coupled to each other via the off-diagonal terms. 
The eigenvalues of the coupled growth rates are 
$\Gamma_\pm=(\Gamma_{\phi,\phi}+\Gamma_{\rm s,s}\pm 
\sqrt{(\Gamma_{\phi,\phi}-\Gamma_{\rm s,s})^2+4\Gamma_{{\rm s},\phi}
\Gamma_{\phi,{\rm s}}})/2$. 
The peak positions indicated in Fig. \ref{fig:spinodal} are 
calculated by solving $d\Gamma_X(q)/dq=0$, where $\Gamma_X(q)$ 
represents $\Gamma_+(q)\,,\Gamma_{\rm t,t}(q)$, and $\Gamma_{\rm b,b}(q)$. 

Shimada {\it et al.} derived similar dynamic equations 
for spinodal decomposition in an LCP solution \cite{Shimada1988c} with the same initial equations; however, they derived their equations by expanding 
the Fokker-Planck equation directly and 
we obtain the free-energy functional and dynamic equations separately. 
However, it is noted that these differences are insignificant in the 
essential features of our results.

\end{document}